\documentclass[showpacs,aps,amssymb,floatfix,prd,amsmath,preprintnumbers]{revtex4}
\usepackage{epstopdf}
\usepackage{capt-of}
\usepackage{graphicx}  
\usepackage{dcolumn}   
\usepackage{bm}
\begin{document}

\title{\bf  Anisotropic cosmological reconstruction in $f(R,T)$ gravity}

\author{B. Mishra \footnote{Department of Mathematics, Birla Institute of Technology and Science-Pilani, Hyderabad Campus, Hyderabad-500078, India, E-mail:bivudutta@yahoo.com },Sankarsan Tarai \footnote{Department of Mathematics, Birla Institute of Technology and Science-Pilani, Hyderabad Campus, Hyderabad-500078, India, E-mail:tsankarsan87@gmail.com},  S. K. Tripathy\footnote{Department of Physics, Indira Gandhi Institute of Technology, Sarang, Dhenkanal, Odisha-759146, India, E-mail:tripathy\_ sunil@rediffmail.com}
}

\affiliation{ }

\begin{abstract}
Anisotropic cosmological models are constructed in $f(R,T)$ gravity theory to investigate the dynamics of universe concerning the late time cosmic acceleration. Using a more general and simple approach, the effect of the coupling constant and anisotropy on the cosmic dynamics have been investigated. In the present study it is found that cosmic anisotropy affects substantially the cosmic dynamics. 
\end{abstract}

\maketitle

\section{Introduction} 
Cosmological models are constructed in recent times to account for the predicted late time cosmic acceleration usually by incorporating possible dark energy candidates in the field equations or by modifying the geometrical part of the action. Amidst  the debate that, whether dark energy exists or whether there really occurs a substantial cosmic acceleration \citep{Riess98, Perlm99, Nielsen16}, researchers have devoted a lot of time in proposing different dark energy models. These models are also tested against the observational data accumulated over a long period of time. Some vector-tensor models are also proposed to explain the cosmic speed up phenomena without adopting these approaches. In these vector-tensor models, the presence of a vector field such as the electromagnetic field provides the necessary acceleration \citep{Fer07, Jim08,Jim09, Jim09a, Dale12}. Usually in General Relativity (GR), it is not possible to explain the late time cosmic acceleration without the assumption of additional dynamical degrees of freedom besides the tensor modes. Some scalar fields are considered as a solution to this. These scalar fields are usually ghost fields having negative kinetic energy, at least around flat, cosmological or spherically symmetric backgrounds e.g. Bulware-Desser mode in massive gravity \citep{Bul72}, bending mode in the self-accelerating branch of Dvali-Gabadadze-Porrati model \citep{Koyama07, Sbisa15, Gumru16}.  Among all the constructed models to understand the cosmic speed up phenomena, geometrically modified gravity theories have attracted substantial research attention. In geometrically modified theories, instead of incorporating some additional matter fields (may be ghost scalar fields), the Einstein Hilbert action is modified considering some extra geometrical objects. These models thereby provide a ghost free and stable alternative to GR. In this context, Harko et al. have proposed $f(R,T)$ gravity theory in which, the geometry part of the action has been modified in such a manner that, the usual Ricci Scalar $R$ in the action is replaced by a function $f(R,T)$ of Ricci Scalar $R$ and the trace of the energy momentum tensor $T$ \citep{Harko11}. In that work, Harko et al. have suggested three different possible forms of the functional $f(R,T)$ such as $f(R,T)= R+2f(T)$, $f(R,T)= f_1(R)+f_2(T)$ and $f(R,T)= g_1(R)+g_2(R)g_3(T)$, where $f(T), f_1(R), f_2(T), g_1(R), g_2(R)$ and $g_3(T)$ are some arbitrary plausible functions or $R$ and $T$.  In $f(R,T)$ gravity theory, the cosmic acceleration is achieved from the geometrical modification and a bit of matter content coupled to the geometrical part of the action. Many workers have used different forms of these functionals to address the issue of mysterious dark energy and the late time cosmic phenomena \cite{Shami15, Myrza12, Hound12, Alvar13, Barri14, Vacar14, Shari12, Shari14, Yous16a, Yous16b, Sahoo16, Aktas17}. Alvarenga et al. studied the scalar perturbations \citep{Alvar13}, Shabani and Ziaie studied the stability of the model \citep{Shab17, Shab17a}, Sharif and Zubair investigated the energy conditions and stability of power law solutions \citep{Sharif13} in this modified gravity theory. Sharif and Zubair \citep{Sharif12a} and Jamil et al. \citep{Jamil12} have studied thermodynamic aspects of $f(R,T)$ theory. There are some good works available in literature in the context of astrophysical applications of this theory  \citep{Moraes17, Zubair15, Alha16}.

With the advent of recent observations regarding the cosmic anisotropy \citep{Anton10, Planck14, Javan15, Lin16, Bengaly15, Zhou17, Andrade18}, there has been an increase in the interest to investigate on the breakdown of the standard cosmology based on cosmic isotropy \citep{Campa06, Grup10, SKT14, Saadeh16, SKT17, Deng18}. In view of this, anisotropic cosmological models that bear a similarity to Bianchi morphology have gained much  importance \citep{Koivisto08, SKT15, Campa07, Campa09}. In the context of geometry modification to explain the late time cosmic dynamics and to take into account the cosmic anisotropy, many workers have constructed some Bianchi type cosmological models in $f(R,T)$ gravity \cite{ Mishr14, Mishr16, Sharif17, Mishra18}. However, a lot remain to be explored in this modified gravity theory in the context of different unanswered issues concerning the late time cosmic acceleration and cosmic anisotropy.\\

In the discussion of cosmological models, space-times admit a three-parameter group of automorphisms are important. When the group is simply transitive over three-dimension, constant-time subspace is useful. Bianchi has shown that there are only nine distinct sets of structure constants for groups of this type. So, Bianchi type space-times admit a three parameter group of motions with a manageable number of degrees of freedom. Kramer et al.\cite{Kramer80} provided a complete list of Bianchi types I–IX space-times. In this work, we have constructed some anisotropic cosmological models in $f(R,T)$ gravity. We have adopted a simple approach to the cosmic anisotropy to investigate the effect of anisotropy on cosmic anisotropy. In order to provide some anisotropic directional pressure, we have considered an anisotropic source along x-direction such as the presence of one dimensional cosmic strings. The effect of the coupling constant in the determination of the cosmic evolution has been investigated. We organise the work as follows: In Sect-II, some basic equations concerning different properties of the universe are derived for Bianchi $VI_h$ model in the framework of the modified $f(R,T)$ gravity. The dynamical features of the models are discussed in Sect-III. Considering the dominance of quark matter that have not yielded to the hadronization process, we have derived the quark energy density and pressure and their evolutionary behaviour in Sect-IV. We conclude in Sect-V.

\section{Basic Equations}

The field equation in $f(R,T)$ gravity for the choice of the functional $f(R,T)=f(R)+f(T)$ is given by \cite{Harko11, Mishra16a}
\begin{equation} \label{eq:2}
f_R (R) R_{ij}-\frac{1}{2}f(R)g_{ij}-\left(\nabla_i \nabla_j-g_{ij}\Box\right)f_R(R)=\left[8\pi +f_T(T),\right]T_{ij}+\left[f_T(T)p+\frac{1}{2}f(T)\right]g_{ij}
\end{equation}
where $f_R=\frac{\partial f(R)}{\partial R}$ and $f_T=\frac{\partial f(T)}{\partial T}$. We wish to consider a functional form of $f(R,T)$ so that the field equations in the modified gravity theory can be reduced to the usual field equations in GR under suitable substitution of model parameters. In this context, we have a popular choice, $f(R,T)=R+2\beta T$ \citep{Das17, Mishr14, Mishr16, Moraes15, Shamir15}. However, we  consider a time independent cosmological constant $\Lambda_0$ in the functional so that $f(R,T)= R+2\Lambda_0+2\beta T$. Here $\beta$ is a coupling constant. For this particular choice of the functional $f(R,T)$, the field equation in the modified theory of gravity becomes,

\begin{equation} \label{eq:3}
R_{ij}-\frac{1}{2}Rg_{ij}=\left[8\pi+2\beta\right]T_{ij} + \left[\left(2p+T\right)\beta+\Lambda_0\right] g_{ij}
\end{equation}
which can also be written as
\begin{equation} \label{eq:3}
R_{ij}-\frac{1}{2}Rg_{ij}=\left[8\pi+2\beta\right]T_{ij} + \Lambda(T) g_{ij}.
\end{equation}

Here $\Lambda(T)=\left(2p+T\right)\beta+\Lambda_0$ can be identified as the effective time dependent cosmological constant. If $\beta=0$, the above modified field equation reduces to the Einstein field equation in GR with a cosmological constant $\Lambda_0$. One can note that, the effective cosmological constant $\Lambda(T)$ picks up its time dependence through the matter field. For a given matter field described through an energy momentum tensor, the effective cosmological constant can be expressed in terms of the matter components. In the present work, we consider the energy momentum tensor as $T_{ij}=(p+\rho)u_iu_j - pg_{ij}-\xi x_ix_j$, where $u^{i}u_{i}=-x^{i}x_{i}=1$ and $u^{i}x_{i}=0$. In a co moving coordinate system, $u^{i}$ is the four velocity vector and $p$ is the proper isotropic pressure of the fluid. $\rho$ is the energy density and $\xi$ is the string tension density.  The strings are considered to be one dimensional and thereby contribute to the anisotropic nature of the cosmic fluid. The direction of the cosmic strings is represented through $x^{i}$ that are orthogonal to $u^{i}$. 

The field equations \eqref{eq:3} of the modified $f(R,T)$ gravity theory, for Bianchi type $VI_h$ space-time  described through the metric $ds^2 = dt^2 - A^2dx^2- B^2e^{2x}dy^2 - C^2e^{2hx}dz^2$ now have the explicit forms 

\begin{equation} \label{eq:6}
\frac{\ddot{B}}{B}+\frac{\ddot{C}}{C}+\frac{\dot{B}\dot{C}}{BC}- \frac{h}{A^2}= -\alpha(p-\xi) +\rho \beta+\Lambda_0   
\end{equation}
\begin{equation} \label{eq:7}
\frac{\ddot{A}}{A}+\frac{\ddot{C}}{C}+\frac{\dot{A}\dot{C}}{AC}- \frac{h^2}{A^2}=-\alpha p +(\rho+\xi)\beta+\Lambda_0   
\end{equation}
\begin{equation} \label{eq:8}
\frac{\ddot{A}}{A}+\frac{\ddot{B}}{B}+\frac{\dot{A}\dot{B}}{AB}- \frac{1}{A^2}=-\alpha p +(\rho+\xi)\beta+\Lambda_0   
\end{equation}
\begin{equation} \label{eq:9}
\frac{\dot{A}\dot{B}}{AB}+\frac{\dot{B}\dot{C}}{BC}+\frac{\dot{C}\dot{A}}{CA}-\frac{1+h+h^2}{A^2}=
\alpha \rho -\left(p-\xi\right)\beta +\Lambda_0   
\end{equation}
\begin{equation} \label{eq:10}
\frac{\dot{B}}{B}+ h\frac{\dot{C}}{C}- (1+h)\frac{\dot{A}}{A}=0.
\end{equation} 

An over dot over a field variable denotes ordinary differentiation with respect to the cosmic time. Here $\alpha = 8\pi+3\beta$ and $A=A(t),B=B(t), C=C(t)$.  An interesting component in this space time is the constant exponent $h$, which takes integral values $-1,0,1$. These three integral values decide the behaviour of the model. However, Tripathy et al. \cite{skt15} and Mishra et al. \cite{skt16} have shown from the calculation of the energy and momentum of diagonal Bianchi type universes that, the value $h=-1$ is favoured compared to other values. Moreover, only in this value of the exponent $h$, the total energy of an isolated universe vanishes.  In view of this, in the present work, we assume this value of $h$ i.e. $h=-1$ and study the dynamics of the anisotropic universe in presence of anisotropic energy sources. The directional Hubble rates may be considered as $H_x=\frac{\dot{A}}{A}$, $H_y=\frac{\dot{B}}{B}$ and $H_z=\frac{\dot{C}}{C}$. With $h=-1$, it is straightforward to get $H_y=H_z$ from \eqref{eq:10} and consequently the mean Hubble parameter becomes, $H=\frac{1}{3}(H_x+2H_z)$. The set of field equations can be reduced to

\begin{eqnarray} 
2\dot{H_z}+3H^2_z+\frac{1}{A^2} &=& -\alpha(p-\xi)+\rho\beta+\Lambda_0,\label{eq:16}\\
\dot{H_x}+\dot{H_z}+H^2_x+H^2_z+H_xH_z-\frac{1}{A^2} &=& -\alpha p+\left(\rho+\xi\right)\beta+\Lambda_0,\label{eq:17}\\
2H_xH_z+H_z^2-\frac{1}{A^2} &=& \alpha \rho-\left(p-\xi\right)\beta+\Lambda_0. \label{eq:18}
\end{eqnarray}

From the above field equations \eqref{eq:16}-\eqref{eq:18}, we obtain the expressions for pressure, energy density and the string tension density as

\begin{eqnarray}
p &=& \frac{1}{\alpha^2-\beta^2}\left[\left(s_1-s_2+s_3\right)\beta-s_2\alpha+\left(\alpha-\beta\right)\Lambda_0\right],\label{eq:19}\\
\rho &=& \frac{1}{\alpha^2-\beta^2}\left[s_3\alpha-s_1\beta-\left(\alpha-\beta\right)\Lambda_0\right], \label{eq:20}\\
\xi &=& \frac{s_1-s_2}{\alpha-\beta}. \label{eq:21}
\end{eqnarray}

Consequently, the equation of state parameter $\omega$ and the effective cosmological constant $\Lambda$ can be expressed as
\begin{eqnarray}
\omega &=& -1+\left(\alpha+\beta\right)\frac{s_2-s_3}{s_1\beta-s_3\alpha+\left(\alpha-\beta\right)\Lambda_0},\label{eq:22}\\
\Lambda &=& \frac{\beta}{\alpha^2-\beta^2}\left[(s_2+s_3)\alpha-(2s_1-s_2+s_3)\beta-(\alpha+\beta)(s_2-s_1)-2(\alpha-\beta)\Lambda_0\right]+\Lambda_0.\label{eq:23}
\end{eqnarray}

In the above equations, $s_1, s_2$ and $s_3$ are functions of the directional Hubble parameters and scale factor: $s_1=2\dot{H_z}+3H^2_z+\frac{1}{A^2}$, $s_2=\dot{H_x}+\dot{H_z}+H^2_x+H^2_z+H_xH_z-\frac{1}{A^2}$ and $2H_xH_z+H_z^2-\frac{1}{A^2}$. Eqns \eqref{eq:19}-\eqref{eq:23} describe the dynamical behaviour of the model. Once the evolutionary behaviour of the functions $s_1, s_2$ and $s_3$ are obtained from some assumed dynamics, the dynamical nature of the model can be studied easily and the modified gravity model can be reconstructed accordingly.

\section{Dynamical Parameters}
We intend to investigate the cosmic history through the assumption of an assumed dynamics concerning the late time cosmic acceleration. In view of this, we assume the scalar expansion be governed by an inverse function of cosmic time i.e. $\theta=(H_x+2H_z)=\frac{m}{t}$ and also we assume that $\theta$ be proportional to the shear scalar $\sigma^2=\frac{1}{2}\left(\sum H_i^2-\frac{1}{3}\theta^2\right); i=x, y,z$. Consequently, $H_x=\left(\frac{km}{k+2}\right)\frac{1}{t}$, $H_y=H_z=\left(\frac{m}{k+2}\right)\frac{1}{t}$. The directional scale factors can be expressed as $A=t^{km/(k+2)}$, $B=C=t^{m/(k+2)}$.

For such an assumption, the functions $s_1, s_2$ and $s_3$ reduce to 
\begin{eqnarray}
s_1 &=& \left[\frac{3m^2-2(k+2)m}{(k+2)^2}\right]\frac{1}{t^2}+\frac{1}{t^{\frac{2km}{k+2}}}, \\
s_2 &=& \left[\frac{(k^2+k+1)m^2-(k+1)(k+2)m}{(k+2)^2}\right]\frac{1}{t^2}-\frac{1}{t^{\frac{2km}{k+2}}}, \\
s_3 &=& \left[\frac{(2k+1)m^2}{(k+2)^2}\right]\frac{1}{t^2}-\frac{1}{t^{\frac{2km}{k+2}}}.
\end{eqnarray}

 From the field eqns. \eqref{eq:16}-\eqref{eq:18}, the pressure, energy density  and string tension density can be obtained as:

\begin{eqnarray}
p &=& \frac{1}{(\alpha^2-\beta^2)}\left[\left(\frac{\phi_1}{(k+2)^2}\right)\frac{1}{t^2}+\frac{(\alpha+\beta)}{t^{\frac{2km}{k+2}}}+(\alpha-\beta)\Lambda_0\right], \\
\rho &=& \frac{1}{(\alpha^2-\beta^2)}\left[\left(\frac{\phi_2}{(k+2)^2}\right)\frac{1}{t^2}+\frac{(\beta-\alpha)}{t^{\frac{2km}{k+2}}}-(\alpha-\beta)\Lambda_0\right],\\
\xi &=& \frac{1}{(\alpha-\beta)}\left[\frac{(k-1)(m^2-m)}{(k+2)^2t^2}-\frac{2}{t^{\frac{2km}{k+2}}}\right],
\end{eqnarray}
where $\phi_1=m\{(k^2+k-2)\beta+(k^2+3k+2)\alpha\}-m^2\{(k^2-k-3)\beta-(k^2+k+1)\alpha\}$ and $\phi_2=(2k+1)m^2\alpha-(3m^2-2km-4m)\beta$ are redefined constants. These physical quantities evolve with the cosmic expansion. Their evolution is governed  by two time dependent factors: one behaving like $t^{-2}$ and the other behaving as $t^{-\frac{2km}{k+2}}$. Since $m$ and $k$ are positive quantities, the magnitude of the physical  quantities (neglecting their sign) decrease monotonically with cosmic time. It is interesting to note that, $\xi$ also decreases from a large value in the initial epoch to small values at late phase of cosmic evolution. This behaviour of $\xi$ implies that, at the initial phase, more anisotropic components are required than at late phase.

\begin{figure}[h]
\begin{center}
\includegraphics[width=0.8\textwidth]{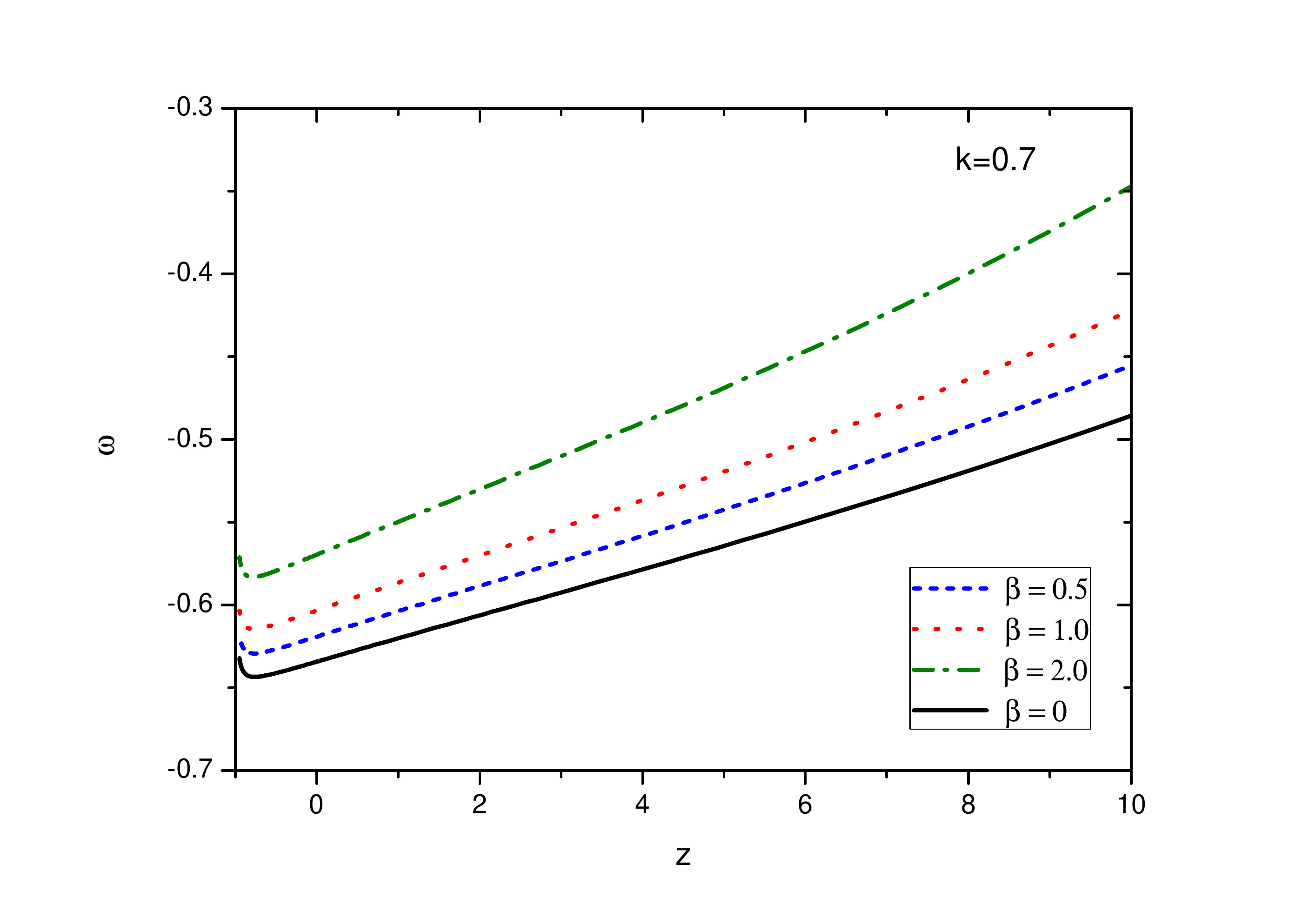}
\caption{Dynamical evolution of the equation of state parameter for different representative values of the coupling constant $\beta$.}
\end{center}
\end{figure}

From eqs. \eqref{eq:19}- \eqref{eq:20}, we obtain the equation of state parameter $\omega=\frac{p}{\rho}$ and the effective cosmological constant $\Lambda$ respectively as

\begin{eqnarray}
\omega &=& -1+(\alpha+\beta)\left[\frac{\phi_3}{\phi_4+(\alpha-\beta)(k+2)^2 \left\{\Lambda_0 t^2-t^{2\left(\frac{k-km+2}{k+2}\right)}\right\}}\right],\\
\Lambda &=&\frac{\beta}{(\alpha^2-\beta^2)} \left[ \frac{\phi_5}{(k+2)^2t^2}-\frac{2(\alpha+\beta)}{t^{\frac{2km}{k+2}}}-2(\alpha-\beta)\Lambda_0 \right]-\frac{\phi_6}{(k+2)^2t^2}+\frac{\beta}{(\alpha-\beta)t^{\frac{2km}{k+2}}}+\Lambda_0,
\end{eqnarray}
  
where $\phi_3=(k^2-2k)m^2-(k^2+2k+3)m$, $\phi_4=(3m^2-2km-4)\beta-(2k+1)m^2 \alpha$,  $\phi_5=\{(k+1)\alpha+(k-3)\beta\}(m^2-m)$ and $\phi_6=\frac{\beta(k-1)(m^2-m)}{(\alpha-\beta)}$ are some constants.

\begin{figure}[h]
\begin{center}
\includegraphics[width=0.8\textwidth]{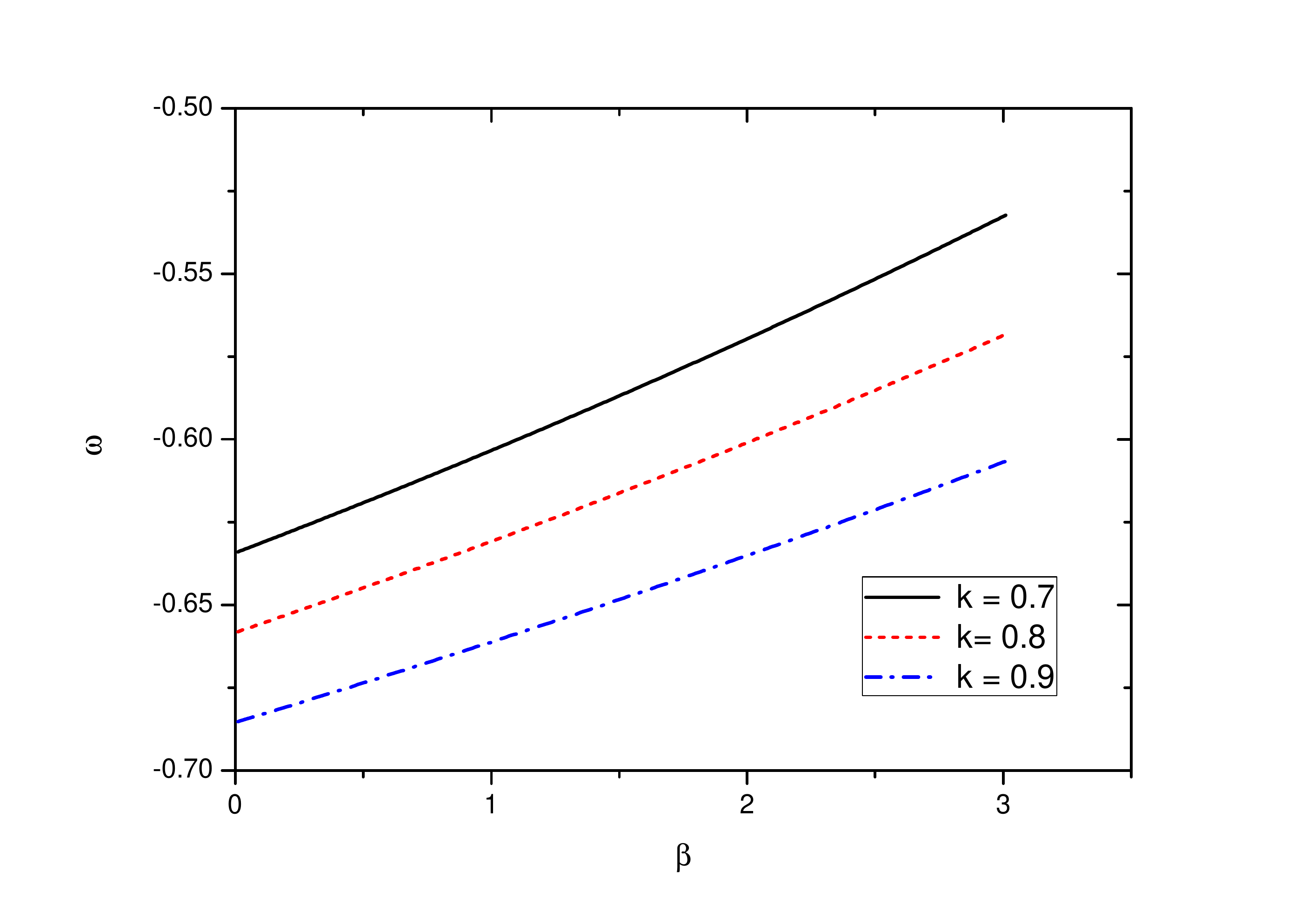}
\caption{Equation of state parameter as function of the coupling constant at present epoch for a given anisotropic parameter.}
\end{center}
\end{figure}

\begin{figure}[h]
\begin{center}
\includegraphics[width=0.8\textwidth]{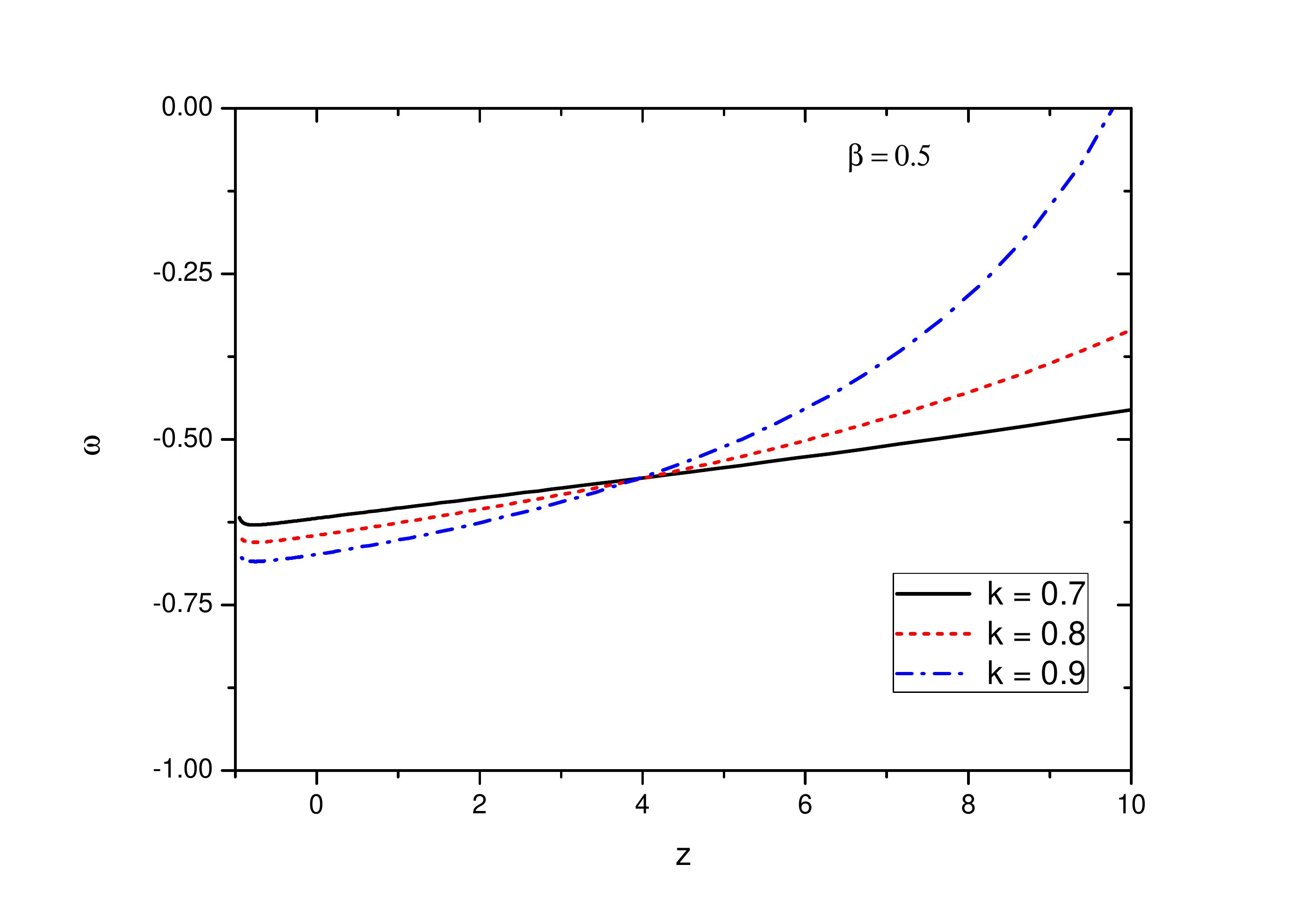}
\caption{Effect of anisotropic parameter on the equation of state.}
\end{center}
\end{figure}

The dynamical nature of the model can be assessed through the evolution of the equation of state parameter  $\omega$. In Figure 1, $\omega$ is plotted as function of redshift for four different values of the coupling constant $\beta$ namely $\beta =0, 0.5, 1.0$ and 2.0. $\beta=0$ refers to the case in GR. The anisotropic parameter is considered to be $k=0.7$ and $m$ is fixed from the observationally constrained value of deceleration parameter $q=-0.598$ \citep{Montiel14}. For all the cases considered here, $\omega$ becomes a negative quantity and remains in the quintessence region through out the period of evolution considered in the work. It decreases from some higher value at the beginning to low values at late times. However, at late phase of cosmic evolution, $\omega$ grows up a little bit which may be due to the anisotropic effect of cosmic strings. 

The coupling constant $\beta$ affects the dynamical behaviour of the equation of state parameter. In order to understand the effect of the $\beta$ on $\omega$, the equation of state at the present epoch is plotted as a function of $\beta$ in Figure 2 for three different values of $k$. One can note that, $\omega$ increases with the increase in the value of the coupling constant. In view of the recent observations predicting an accelerating universe, the value of coupling constant $\beta$ should have a lower value i.e. $\beta \leq 1$. 

In Figure 3, we have shown the effect of anisotropy on the equation of state parameter. In the figure, we assume three representative values of the anisotropy i.e $k=0.7, 0.8$ and 0.9 for a given coupling constant $\beta=0.5$. Anisotropy brings a substantial change in the magnitude as well as the behaviour of the equation of state parameter. There occurs a flipping behaviour of $\omega$ at a redshift  $z_f \simeq 4$. At a cosmic time earlier to $z_f$, with the increase in the anisotropy of the model, $\omega$ assumes a higher value. In other words, prior to $z_f$, higher the value of $k$, higher is the $\omega$. It displays  an opposite behaviour at cosmic times later to $z_f$. Also, at the redshift $z_f$, curves corresponding to all $k$ considered here cross each other. In general, the rate of evolution of the equation of state parameter increases with the increase in the value of the anisotropic parameter. 
\section{Anisotropic universe with quark matter}
One can believe that, quarks and gluons did not yield to hadronization and resisted as a perfect fluid that spread over the universe and may contribute to the accelerated expansion. Here we will reconstruct an anisotropic cosmological model with non interacting quarks that may well be dealt as a Fermi gas with an  equation of state given by \cite{Kapusta94, Aktas07}

\begin{equation}
p_q=\frac{\rho_q}{3}-B_c,
\end{equation}
where $p_q$ is the quark pressure, $\rho_q$ is the quark energy density and $B_c$ is the bag constant. We assume that quarks exist along with one dimensional cosmic strings without any interaction.  The quark energy density can then be expressed as $\rho_q=\rho-\xi-B_c$. Going in the same manner as described in the previous section, we can have the expressions for the quark pressure and quark energy density as
\begin{eqnarray}
\rho_q &=& \frac{1}{\alpha^2-\beta^2}\left[(\alpha+\beta)s_2+s_3\alpha-(\alpha+2\beta)s_1-\left(\alpha-\beta\right)\Lambda_0\right]-B_c, \label{eq:26}\\
p_q &=& \frac{1}{3\left(\alpha^2-\beta^2\right)}\left[(\alpha+\beta)s_2+s_3\alpha-(\alpha+2\beta)s_1-\left(\alpha-\beta\right)\Lambda_0\right]-\frac{4B_c}{3} \label{eq:27}
\end{eqnarray}

If we put $\beta=0$, the model reduces to that in GR with a cosmological constant. In that case, the above equations reduce to
\begin{eqnarray}
\rho_q &=& \frac{1}{8\pi}\left[s_2+s_3-s_1-\Lambda_0\right]-B_c, \label{eq:28}\\
p_q &=& \frac{1}{24\pi}\left[s_2+s_3-s_1-\Lambda_0\right]-\frac{4B_c}{3} \label{eq:29}
\end{eqnarray}

Substituting the expressions for $s_1, s_2$ and $s_3$ in eqs. \eqref{eq:26} and \eqref{eq:27}, the quark matter energy density and quark pressure are obtained as

\begin{eqnarray}
\rho_q &=& \frac{1}{\alpha^2-\beta^2}\left[\frac{\phi_7}{(k+2)^2}\frac{1}{t^2}+\frac{(\alpha+3\beta)}{t^{\frac{2km}{k+2}}}-\left(\alpha-\beta\right)\Lambda_0\right]-B_c,\\
p_q &=& \frac{1}{3(\alpha^2-\beta^2)}\left[\frac{\phi_7}{(k+2)^2}\frac{1}{t^2}+\frac{(\alpha+3\beta)}{t^{\frac{2km}{k+2}}}-\left(\alpha-\beta\right)\Lambda_0\right]-\frac{4B_c}{3},
\end{eqnarray}
where $\phi_7=\phi_2-(k-1)(m^2-m)(\alpha+\beta)$. For some reasonable value of the coupling parameter $\beta$ and the anisotropic parameter $k$, the quark energy density and quark pressure decrease smoothly with the cosmic evolution. Bag constant certainly has a role to play at late times when the value of $\rho_q$ and $p_q$ are mostly dominated by this quantity.
\section{Conclusion}

This paper reports the investigation of the dynamical behaviour of an anisotropic Bianchi type $VI_h$ universe in the presence of one dimensional cosmic strings and quark matter. Anisotropic cosmological models are reconstructed for a power assumption of the scale factor in the frame work of $f(R,T)$ gravity.   In the process of reconstruction and study of dynamical features of the model, we chose the functional $f(R,T)$ as $f(R,T)=R+2\Lambda_0+2\beta T$. From some general expressions of the physical quantities, we derived the expression of the equation of state parameter and the effective cosmological constant. The effects of anisotropy $k$ and the coupling constant $\beta$ are investigated. It is observed that, with an increase in the coupling constant the equation of state parameter assumes a higher value. Anisotropy is observed to affect largely to the dynamics of the model. The equation of state parameter undergoes an increased rate of growth with an increase in the anisotropy. We hope, the present study will definitely put some light in the context of the uncertainty prevailing in the studies of the late time cosmic phenomena.

\section{Acknowledgment}
BM and SKT thank IUCAA, Pune (India) for hospitality and support during an academic visit where a part of this work is accomplished.ST thanks University Grants Commission (UGC), New Delhi, India, for the financial support to carry out the research work. The authors are very thankful to the anonymous reviewer for his useful comments that helped us to improvise the manuscript.

\end{document}